# END CONDITIONS OF PIANO STRINGS

Kerem Ege[1], Antoine Chaigne[2]

[1]*Laboratory for Solid Mechanics, Ecole Polytechnique, UMR7649,
91128 Palaiseau Cedex, kerem.ege@lms.polytechnique.fr*
[2]*Unité de Mécanique, Ecole Nationale Supérieure de Techniques Avancées,
Chemin de la Hunière, 91761 PALAISEAU Cedex, antoine.chaigne@ensta.fr*

**Abstract**

The end conditions of piano strings can be approximated by the input admittance at the bridge. Proper measurements of this value are therefore required. A method of validation of admittance measurements on simple structures is proposed in this paper. High resolution signal analysis performed on string's vibrations yields an estimate for the input admittance. This method is implemented on a simplified device composed of a piano string coupled to a thin steel beam.

## INTRODUCTION

The research of a trade-off between *loudness* and *sustain* (duration) is a major issue for piano designers and manufacturers. The way the energy of vibration is transferred from the piano string to the soundboard depends on the end conditions of the strings at the bridge: these conditions can be approximated by the *input admittance* at the connecting point between the string and the resonator. Therefore, proper measurements of this value are needed.

Given this, we propose here a method of validation of admittance measurements on simple structures. Parameters such as frequencies and damping factors of the string partials depend directly on the end conditions. The analysis of the vibratory signal of the string, based on high resolution estimation methods (ESPRIT algorithm), allows us to evaluate efficiently those parameters and leads to the calculation of the input admittance.

This method is implemented on a simplified device composed of a piano string coupled to a thin steel beam. The comparative study of two experimental cases (*isolated* string vs. *coupled* string) leads us to the input admittance. This value, derived from vibratory measurements of the string is compared to direct admittance measurements performed on the beam, and to theoretical predictions, in order to validate the method.

## STRING-SOUNDBOARD COUPLING IN PIANOS

### Summary of piano acoustics

The vibrations of the piano string are coupled to the soundboard (which radiates the sound) via the bridge (Figure 1). This coupling determines the tone duration and the sound power. The boundary conditions of the string must ensure a compromise between sound power efficiency (high transverse velocity for a given string force) and tone duration (low soundboard velocity).

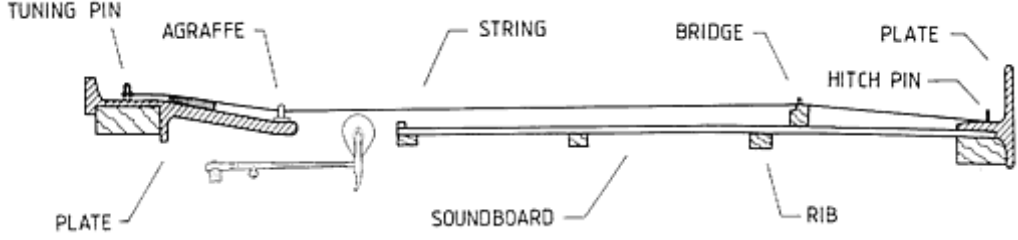

Figure 1: Principal sketch of the piano,
with the main components. [Askenfelt, 1990]

**Review of the linear model**

Let us consider the bridge velocity $v(L,t) = \frac{\partial \xi}{\partial t}(L,t)$ (where $\xi(x,t)$ is the vertical transverse displacement of the string), and the force transmitted to the bridge $f(L,t) = -T\frac{\partial \xi}{\partial x}(L,t)$ (where T is the tension of the string). Figure 2 shows these vectors at the end of the string.

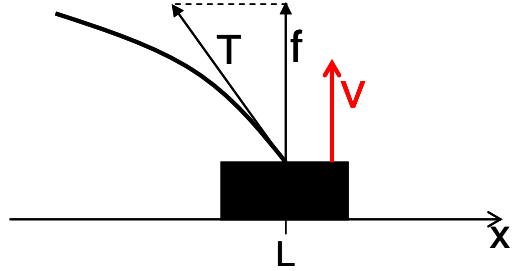

Figure 2: Force and velocity at the bridge

A proper model of this coupling leads us to consider the mean mechanical power transmitted from the string to the soundboard for a steady state excitation of the soundboard at angular frequency $\omega$ :

$$P_m(\omega) = \frac{1}{2}\text{Re}\left\{F(L,\omega)V^*(L,\omega)\right\}$$

The input admittance (at the bridge side) is the ratio between the applied force and the velocity of the soundboard at point $x = L$ (see figure 2) :

$$Y(\omega) = \frac{V(L,\omega)}{F(L,\omega)} = G(\omega) + jB(\omega) . \qquad (1)$$

We have therefore :

$$P_m(\omega) = \frac{1}{2}\text{Re}\left\{F(L,\omega)F^*(L,\omega)Y(\omega)\right\} = \frac{1}{2}G(\omega)Z_c^2|V(L,\omega)|^2 \qquad (2)$$

where $Z_c = \frac{T}{c} = \sqrt{\mu T} = \frac{F(L,\omega)}{V(L,\omega)}$ is the characteristic impedance of the string (at the string side), with $c = \sqrt{\frac{T}{\mu}}$ the celerity of the transverse waves in the string and $\mu$ the linear density of the string.

Proper measurements of Y are therefore needed. Previous measurements of soundboard impedance Z have been done (Figure 3) but need to be reconsidered especially in the upper frequency range.

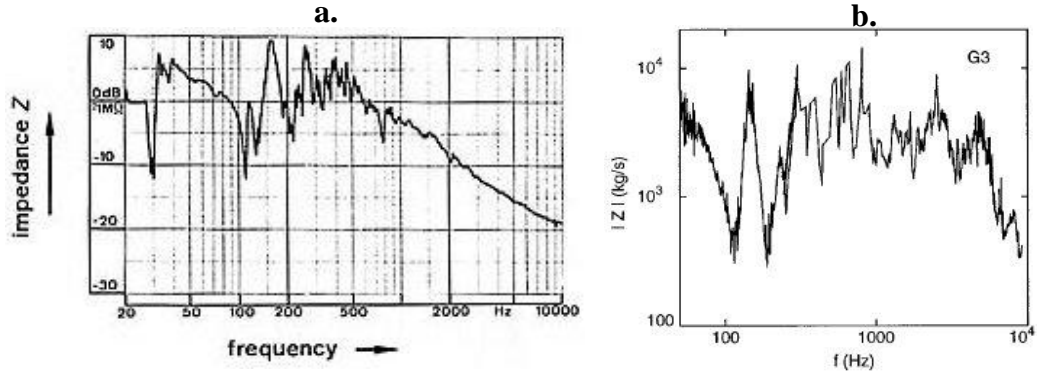

*Figure 3: Previous measurements of soundboard impedance*
*a. [Wogram,1984]      b. [Giordano,1998]*

In particular, the fact that the modulus of the impedance decreases with frequency (above 1 kHz for Wogram's results and above 7kHz for Giordano's ones) seems to contradicts the theory which predicts a constant asymptotical value (see for example [Busch-Vishniac, 1981] ):

$$Z(\omega) \to 8\sqrt{\rho_p h D} \qquad (3)$$

where $\rho_p$ is the density of material, $h$ the thickness, $D = Eh^3/[12(1-\nu^2)]$ the rigidity of an isotropic plate, $E$ the Young's modulus and $\nu$ the Poisson's ratio. Notice that for an orthotropic plate, a similar result applies where $E$ needs to be replaced by $\sqrt{E_1 E_2}$.

From a physical point of view, the fact that the impedance decreases would mean that the mobility of the soundboard increases in the upper frequency range, which is rather questionable.

The soundboard impedance is defined as the inverse of the soundboard admittance. For a single polarisation of the string we have:

$$Z(\omega) = \frac{F(L,\omega)}{V(L,\omega)} = \frac{1}{Y(\omega)} = \frac{G(\omega) - jB(\omega)}{G(\omega)^2 + B(\omega)^2}$$

**String model**

The soundboard is a moving end for the string which modifies its eigenfrequencies $f_n$ and damping factors $\alpha_n$ in $s^{-1}$ (inverse of decay times).

An usual transmission line model yields the admittance of the string at the end:

$$Y(\omega) = -jY_c \tan k_n L \qquad (4)$$

where $Y_c = \dfrac{1}{Z_c}$ is the characteristic admittance of the string and $k_n$ the wave number.

With a first-order approximation $\tan k_n L \approx L\dfrac{\delta\omega_n}{c}$ we get the perturbation of the complex eigenfrequencies:

$$\delta\omega_n = \frac{jcZ_c Y(\omega)}{L} = \frac{j\omega_1 Z_c}{\pi}[G(\omega) + jB(\omega)] = j\alpha_n + 2\pi\,\delta f_n \qquad (5)$$

where $\omega_1 = \dfrac{\pi c}{L}$ is the angular frequency for the first transverse mode of the string.

This finally yields the perturbation of the eigenfrequencies (real part) and to the damping factors (imaginary part):

$$\alpha_n = \frac{T}{L} G(\omega) \text{ and } \delta f_n = \frac{-T}{2\pi L} B(\omega) \quad (6)$$

**The method and its validation**

Since these modifications of string's eigenfrequencies and decay times are directly related to the soundboard admittance, we propose to measure these quantities in order to derive the soundboard admittance. In addition, direct measurement of the admittance at the bridge will allow to check whether the admittance derived from string's measurements is really the one that is "seen" by the string.

The validation of the method is divided in four steps:
(a) Measurements of eigenfrequencies and decay times on an isolated string.
(b) Same measurements with the string loaded at one end by a known admittance.
(c) Derive the endadmittance from these measurements.
(d) Compare the results with calculated or directly measured load admittance.

## EXPERIMENTAL SET-UP

**The prototype**

The experimental device is composed of a piano string stretched out between two ends screwed to a massive support (an aluminum plate). Two cases are considered: the first, Figure 4a, is an isolated string (with two fixed ends assuming to have an infinite admittance) ; the second, Figure 4b, is a string coupled to a thin steel beam (with a known load admittance).

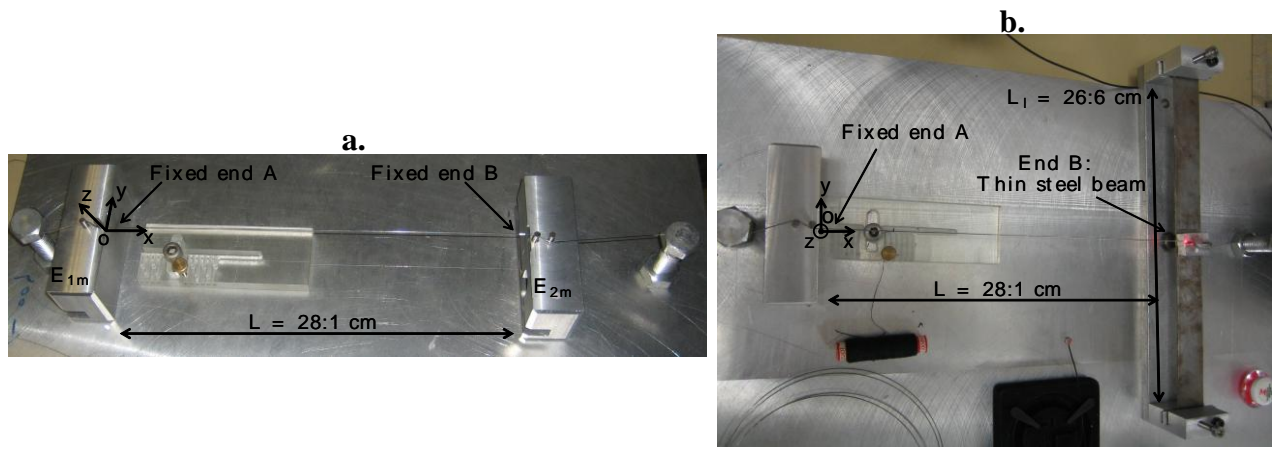

*Figure 4: The prototype*

*a. Isolated string*  *b. String loaded at one end by a known admittance (thin beam)*

**Experimental difficulties**

The main difficulties result from the fact that we have to compare two experimental cases. Fist of all we must keep exactly the same length and the same tension of the string in both experimental cases: with and without the beam. This is critical as long as a small error of one of these two parameters directly affects the eigenfrequencies of the string. Another difficulty is that when attached to the string, the beam is prestressed, which modifies its properties compared to the isolated beam.

An additional source of experimental difficulty lies in the fact that we must avoid exciting the horizontal polarization of the string, otherwise we get double peaks.

Finally, because the frequency shifts due to the load are very small, we have the necessity of using powerful estimation methods (see the next Section).

## RESULTS

### Influence of the string on the load and influence of the load on the string

The system under study is a string coupled to the beam at one end. The aim of the first series of measurements consists in investigating the influence of one component of the system onto the other one. For both the string and beam, we compare measurements done for the uncoupled components (in blue on the figures below) to measurements performed on the coupled system (in red).

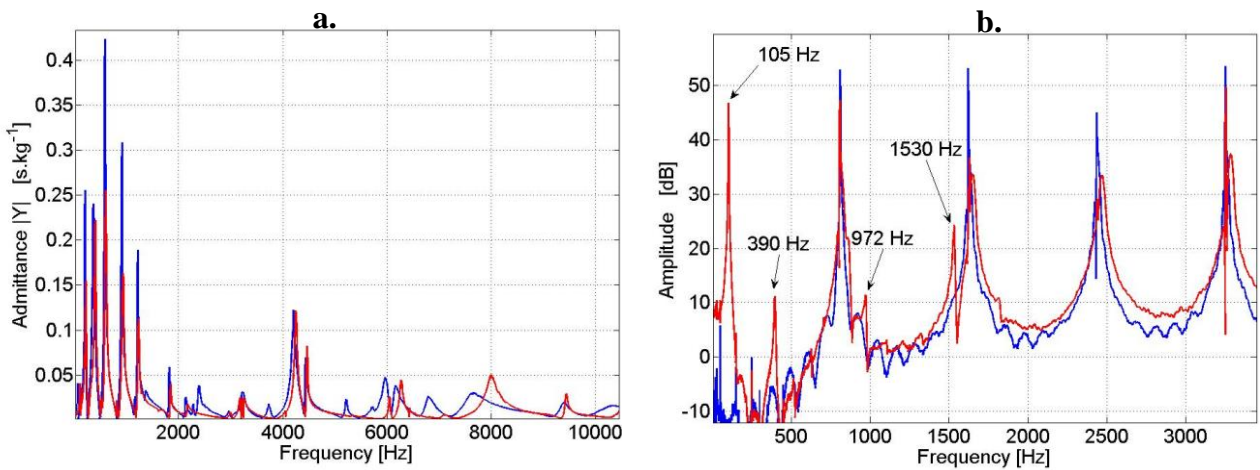

*Figure 5: Influence of the coupling*
*a. Beam Admittance (beam alone in blue, beam coupled to the string in red)*
*b. String's spectrum (below 3.5 kHz) (isolated string in blue, string loaded in red)*

Figure 5a shows the beam admittance at the bridge point for the two cases. These direct measurements were conducted with an impedance head screwed to a mechanical shaker. The comparison of the admittances in both cases shows that the normal flexion modes frequencies have been slightly increased after the coupling. This can be explained by the fact that the prestressing of the beam due to the tension of the string adds stiffness to the system.

The influence of the load on the string is showed in Figure 5b. These spectra are obtained by mean of the Fast Fourier Transform (FFT) applied to the recorded transverse velocity of the string (isolated and coupled) at a given point with the help of a laser vibrometer. Two features can be pointed out: the string's eigenfrequencies are slightly shifted up after the loading and new spectral components appear. The first three components (at 105 Hz, 390 Hz and 972 Hz respectively) are the first flexural modes of the beam; the component at 1530 Hz is a torsional mode of the beam.

## An alternative to Fourier analysis: the ESPRIT algorithm

As it has been underlined earlier (Figure 5b) the frequency shifts due to the load are very small. Therefore, the usual FFT analysis does not appear to be the most appropriate tool for estimating the string's eigenfrequencies and damping factors.

Here, another estimation method has been used: the ESPRIT algorithm. It is a high-resolution estimation method which presents interesting alternatives to classical Fourier transform. This method is based on the assumption that the analyzed signal $x[n]$ sampled at frequency $F_e$ is composed of damped sinusoids (equation 7). This finally yields estimates for the following parameters: frequencies ($f_n = v_n F_e$), damping factors ($\alpha_n = \sigma_n F_e$), amplitudes $A_n$ and phases $\Phi_n$.

$$x[n] = \sum_{i=1}^{L} A_i e^{-\sigma_i n} \cos(2\pi v_i n + \Phi_i) \qquad (7)$$

For more details of this method, we invite the reader to refer to [Roy et al., 1986] and [Roy et al., 1989].

Another method has also been applied: the Hilbert method. This classical method is based on a demodulation technique. Assuming that each components of the signal can be isolated from the others by bandpass filtering (equation 8), a complex signal is associated to the damping sinusoids by means of Hilbert transform (equation 9).

$$x_k[n] = A_k e^{-\sigma_k n} \cos(2\pi v_k n + \Phi_k) \qquad (8)$$

and the associated complex signal is given by:

$$y_k[n] = \mathrm{H}(x_k[n]) = A_k e^{-\sigma_k n} e^{i(2\pi v_k n + \Phi_k)} \qquad (9)$$

Amplitude and phase detection finally leads to the estimations of the sinusoids components through linear regression:

$$\ln |y_k[n]| = \ln A_k - \sigma_k n \qquad (10)$$

$$\arg y_k[n] = 2\pi v_k n + \Phi_k \qquad (11)$$

## Damping factors

Using the analysis tools presented above, we are now able to determine the complex eigenfrequencies of the string. Figure 6 shows the damping factors obtained for the string's partials in both cases of our study. These data are given in the Table below.

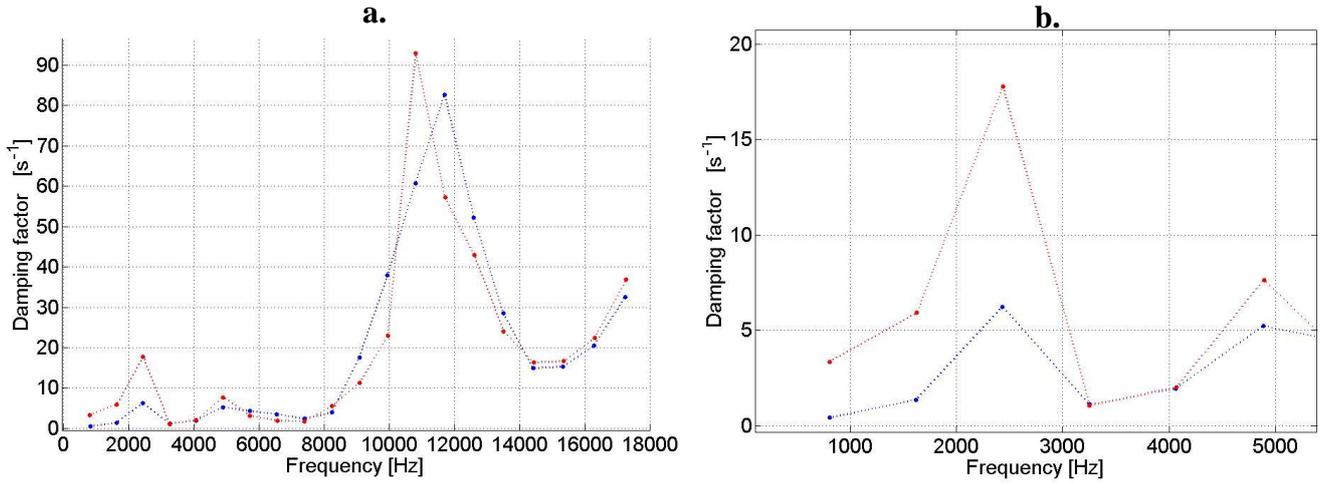

*Figure 6: Damping factors of the string partials (obtain via ESPRIT method)*
*a. Frequency range [0-18 kHz] (isolated string in blue, string loaded in red)*
*b. Zoom-in. Frequency range [0-5.5 kHz] (isolated string in blue, string loaded in red)*

|  | $n = 1$ | $n = 2$ | $n = 3$ | $n = 4$ | $n = 5$ | $n = 6$ |
|---|---|---|---|---|---|---|
| Isolated string $f_n$ (Hz) | 810 | 1621 | 2433 | 3247 | 4064 | 4886 |
| Coupled string $f_n$ (Hz) | 807 | 1628 | 2437 | 3252 | 4070 | 4893 |
| Isolated string $\alpha_n$ ($s^{-1}$) | 0.43 | 1.38 | 6.24 | 1.12 | 1.97 | 5.24 |
| Coupled string $\alpha_n$ ($s^{-1}$) | 3.36 | 5.92 | 17.77 | 1.06 | 2.03 | 7.65 |
| Isolated string $A_n$ | 0.0246 | 0.0221 | 0.0154 | 0.0124 | 0.0227 | 0.0032 |
| Coupled string $A_n$ | 0.0211 | 0.0051 | 0.0043 | 0.0117 | 0.0099 | 0.0041 |

It can be observed in Figure 6b and in the Table that the first partials of the coupled string are more damped than the other ones. Furthermore, the amplitudes are lowered compared to the isolated string. It means that, in this frequency range, there is an energy transfer between string and beam. For the first partial of the string (the fundamental), the decay time $\tau_1 = \dfrac{1}{\alpha_1}$ is only of 0.3 seconds for the coupled string whereas it is equal to 2.3 seconds for the isolated string. Finally, the high damping factors for the upper partials (more than $20\ s^{-1}$ above 10kHz) are mainly due to intrinsic dissipation in the string.

**Load admittance: comparison between direct measurements and calculation**

From measurements of damping factors and frequency shifts, we can derive the load admittance for each complex frequency $\omega_n$:

$$Y(\omega_n) = \frac{-i\pi\,\delta\omega_n}{\omega_1 Z_c} = \frac{\pi\,\alpha_n - i\,2\pi^2\,\delta f_n}{\omega_1 Z_c} \qquad (12)$$

The results, presented in Figure 7, are important. The good agreement (for frequencies smaller than 7 kHz) between direct measurements, theory, and admittance derived from calculation validates the method. These comparisons also confirm that the measured isolated load is the one that is "seen" by the string. Such an approach should be of interest for future measurements on piano soundboard, since, due to the complexity of string-soundboard coupling, one crucial question is to know withoutambiguity whether or not measurements performed on the soundboard itself are of real significance with regard to string's behavior and, in turn, to piano sound.

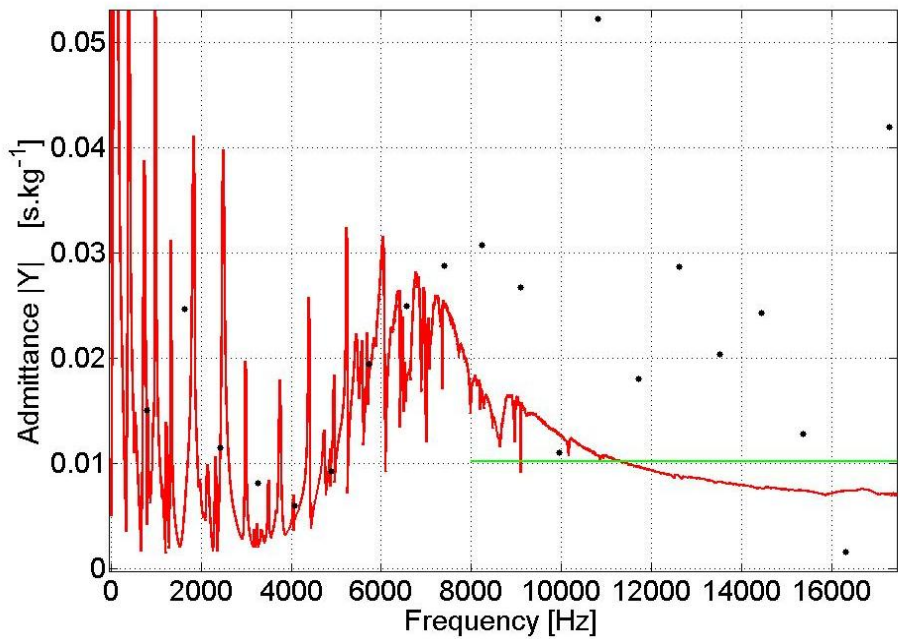

*Figure 7: Load ("soundboard") admittance – Comparison between three methods*
*Direct measurements conducted with a mechanical shaker: in red.*
**Calculation derived from coupled string's spectrum (equation 12): black points**
Theoretical asymptotic value (derived from equation 3): in green

## CONCLUSIONS

In this paper a method for validating admittance measurements on simple coupled structures has been presented. A prototype composed of a single piano string, isolated or coupled to a thin steel beam, has been designed. High resolution signal analysis performed on string's vibrations yields an estimate for the input admittance. This method gives accurate results and should be now applied to measurements on piano strings mounted on real pianos in order to derive soundboard admittance and validate direct measurements.